\documentclass[a4paper]{article}
\usepackage{graphicx, subfigure,multirow,array}
\usepackage{epsfig}
\usepackage{authblk}
\title{Linear and nonlinear viscoelastic arterial wall models: application on animals}
\author[1]{Arthur R. Ghigo}
\author[1]{Xiao-Fei Wang}
\author[2]{Ricardo Armentano}
\author[1]{Pierre-Yves Lagr\'ee}
\author[1]{Jose-Maria Fullana \thanks{jose.fullana@upmc.fr}}
\affil[1]{Sorbonne Universit\'es  UPMC, CNRS UMR 7190, Institut Jean le Rond $\partial$'Alembert}
\affil[2]{Faculty of Engineering and Natural and Exact Sciences, Favaloro University }

\date{Revised version \today, Journal of Biomechanical Engineering.}
\begin{document}

\maketitle

\begin{abstract}
This work deals with the viscoelasticity of the arterial wall and its influence on the pulse waves.
We describe the viscoelasticity by a nonlinear Kelvin-Voigt model in which the coefficients are fitted using experimental time series of pressure and radius measured on a sheep's arterial network.  We obtained a good agreement  between the results of the nonlinear Kelvin-Voigt model and the experimental measurements. We found that the viscoelastic relaxation time - defined by the ratio between the viscoelastic coefficient and the Young's modulus - is nearly constant throughout the network. Therefore, as it is well known that smaller arteries are stiffer, the viscoelastic coefficient rises when approaching the peripheral sites to compensate the rise of the Young's modulus, resulting in a higher damping effect.
We incorporated the fitted viscoelastic coefficients in a nonlinear 1D fluid model to compute the pulse waves in the network. The damping effect of viscoelasticity on the high frequency waves is clear especially at the peripheral sites. 
\end{abstract}

%\tableofcontents

\section{Introduction}
   
One way of obtaining information about the cardiovascular system is by
studying pressure and flow waveforms. By analyzing and modeling flow
waveforms we can deduce the mechanical properties of the
cardiovascular system even in regions of the network inaccessible to
visualization techniques. However, challenges still remain such as
real time observations and analysis of flow waveforms to help the medical
staff make diagnostics and surgical decisions. Pulse waves of pressure
and flow rate in the arterial system can be correctly captured by 1D
models of blood flow. The 1D approach is attractive because it is a
good compromise between modeling complexity and computational cost and
is useful for medical applications such as disease diagnostic and pre-surgical planning.   
Difficulties arise when performing patient-specific simulations as the number of parameters required by the model increases with the number of simulated arterial segments. The hardest parameters to obtain are those describing the complex mechanical properties of the arterial wall.
In the 1D model, a constitutive equation of the wall mechanics is necessary to close the system of conservation laws of mass and momentum.
Although the viscoelastic behavior of the wall has been
recognized as  fundamental for a long time most 1D numerical simulations existing in literature adopted elastic wall models for simplicity since the viscoelastic coefficient is difficult to measure. Another problematic point is that the viscoelastic response of the wall dynamics interacts with the viscoelastic properties of the blood. Therefore both phenomena should be included in the model to obtain a complete picture of the coupled wall-blood flow dynamics even though it was shown that differentiating both behaviors from experimental data is a complicated task \cite{Wang2016}.

Nevertheless, there are previous studies of blood flow in networks using viscoelastic 1D models. The viscoelastic models for the arterial wall fall into roughly two categories: Fung's quasilinear viscoelastic models~\cite{Fung1993biomechanics} and an arrangement of spring-dashpot elements.
Models of the first category are more general but also more difficult to handle when coupled with a 1D model of blood flow because they involve a creep function and convolutions have to be computed~\cite{steele2011predicting}. Holenstein et al.~\cite{holenstein1980viscoelastic} proposed a model and fitted the parameters from published data. Reymond et al.~\cite{reymond2011validation,reymond2009validation} adopted Holenstein's model and parameter values in their patient-specific simulations. Comparison between numerical results and \textit{in vivo} measurements reveals a considerable impact of the viscoelasticity on the pulse waves.
Another comparison between the results obtained with 1D models using different viscoelastic models of the first category shows that the differences between the results computed with different models are minor~\cite{raghu2011comparative}.
Segers et al~\cite{segers1997assessment} proposed another approach incorporating a frequency dependent viscoelastic model with the linearized 1D model 
of blood flow. They found that the influence of viscoelasticity is comparable to the elastic nonlinearity~\cite{raghu2011comparative}. We note  that the parameter values of the viscoelastic model are fitted from limited available data in literature.

The second class of viscoelastic models are build by combinations of springs and dashpots. The Kelvin-Voigt model which consists of one spring and one dashpot connected in parallel is suitable to describe viscoelastic solids and is straightforward to incorporate in a 1D model of blood flow.
Armentano et al.~\cite{armentano1995arterial} fitted the coefficients
of the Kelvin-Voigt model from simultaneous experimental measurements
of diameter and pressure and obtained acceptable agreements with measurements. 
Alastruey et al.~\cite{alastruey2011pulse} also adopted the Kelvin-Voigt model and estimated the parameters using a tensile test.
They simulated the pulsatile flow in an \textit{in vitro} experimental setup and compared this model with an elastic one. They showed that the viscoelastic model agrees much better with measurements than the elastic one. We observe furthermore that the vessels in the study were made of polymers which are actually much less viscous than the real arterial wall.

In this paper, we propose to analyze a nonlinear Kelvin-Voigt wall model and to study the effect of viscoelasticity on the pulse waves of a sheep's arterial networks.
We collected simultaneous time series of diameter and pressure at different arterial sites from a group of sheep (experimental data from \cite{valdez2009analysis}). We estimated the viscoelasticity coefficients by fitting the experimental measurements using the following nonlinear Kelvin-Voigt model
\begin{eqnarray*}
(1 - \eta^2) { R \over h} P = E \epsilon + \phi_0  \dot{\epsilon}  + \phi_{NL} \dot{\epsilon}^2
\label{eq:model}
\end{eqnarray*}
where the nonlinear coefficient $\phi_{NL} $ appears necessary to retrieve the experimental data. In \cite{segers1997assessment} the authors have shown that the nonlinear term in $\epsilon^2$ can be neglected compared to $\dot{\epsilon}$. We have confirmed this fact  and therefore we have removed the corresponding term in the proposed viscoelastic model.  Conversely the nonlinear term in $\dot{\epsilon}^2$ seems to play an important role in the wall dynamics. Reference \cite{erbay1992wave} used this nonlinear term to study the wave propagation in nonlinear viscoelastic tubes, and the theoretical basis of the approach is in \cite{bird1977dynamics}. 

We computed the unsteady blood flow in the network using a 1D blood flow model coupled to the linear viscoelastic wall model. We observed the smoothing effect of the wall viscosity on the pulse waveforms in particular at the terminal sites of the network. This result is corroborated by the observation that the viscoelastic relaxation time $\phi_0 / E$ is nearly constant throughout the network. Therefore, as it is well known that smaller arteries are stiffer, the viscoelastic coefficient rises when approaching the peripheral sites to compensate the increase of the Young's modulus, resulting in a higher damping effect.

Section \ref{sec:metho} presents the experimental protocol for data acquisition, the proposed nonlinear Kelvin-Voigt model, the optimization approach to compute the model parameters, and the 1D blood flow model used in the numerical simulations. In Section \ref{sec:results-discussion} we discuss the optimization results and  the numerical findings. We also present a numerical simulations to explain the differences between an elastic and a viscoelastic wall model.

\section{Methodology}
\label{sec:metho}
\subsection{Data acquisition}
The experimental data were obtained from a group of eleven sheep (male Merino, between 25 and 35 kg).
Before surgeries, the animals were anesthetized with sodium pentobarbital (35 mg/kg).
The arterial segments of interest (6 cm long) were separated from the surrounding tissues.
To measure the diameter, two miniature piezoelectric
crystal transducers (5 MHz, 2 mm in diameter) were sutured on opposite
sides into the arterial adventitia.
The animals were then sacrificed and the arterial segments of interest were excised for \textit{in vitro} tests.

The arterial segments were mounted on a test bench where a periodical flow was generated by an artificial heart (Jarvik Model 5, Kolff Medical Inc., Salt Lake City, USA).  The input signal was close as possible to a physiological waveform. We obtained the desired pressure waveforms by simple adjustments of tuning resistances and Windkessel chambers.

The circulating liquid was an aqueous solution of Tyrode.
At each arterial segment the internal pressure was measured using a solid-state pressure
micro-transducer (Model P2.5, Konigsberg Instruments, Inc., Pasadena, USA), previously calibrated using a mercury manometer at 37$^\circ$C.
The arterial diameter signal was calibrated in millimeters using the 1 mm step
calibration option of the sono-micrometer (Model 120, Triton Technology, San Diego, USA).
The transit time of the ultrasonic signal with a velocity of 1,580 m/s was converted to the vessel diameter. 
The experimental protocol was conformed to the \textit{European
Convention for the Protection of Vertebrate Animals used for
Experimental and Other Scientific Purposes}.
For more details on the animal experiments, please refer to~\cite{valdez2009analysis}.

\begin{figure}[ht]
\centering
\includegraphics[width=\textwidth]{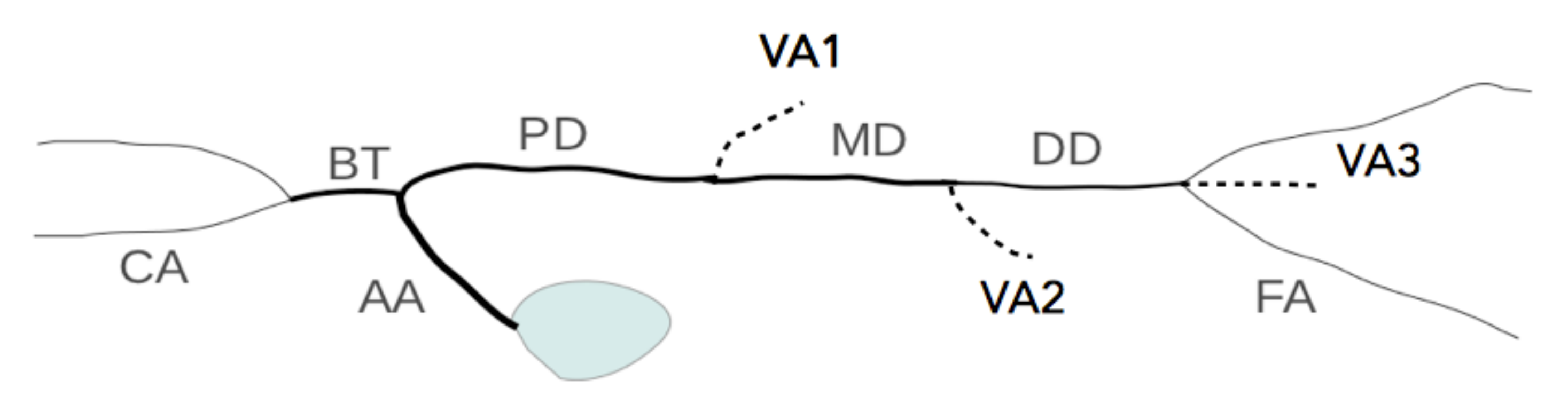}
\caption{Arterial tree of a sheep. Experimental data are collected from eleven
sheep at the following seven locations: Ascending Aorta (AA), Proximal Descending aorta
(PD), Medial Descending aorta (MD), Distal Descending aorta (DD),
Brachiocephalic Trunk (BT), Carotid Artery (CA) and Femoral Artery (FA). 
There are three virtual arteries (VA), which are indicated by dashed lines, to model the side branches
when pulse waves are simulated.
Parameters for all the arteries are shown in Table~\ref{tab:Parameters_sheep_tree}.} 
\label{sheep_tree}
\end{figure}

The synchronized recording of transmural pressure and diameter was
applied on the following seven anatomical locations as shown in Figure~\ref{sheep_tree}:
Ascending Aorta (AA), Proximal Descending aorta
(PD), Medial Descending aorta (MD), Distal Descending aorta (DD),
Brachiocephalic Trunk (BT), Carotid Artery (CA) and Femoral
Artery (FA).

We note that experimental data was acquired from blood vessels that
are extracted from their surrounding tissue and this modifies the the
experimental "pressure-radius" function we will use to do the optimization process.
Others factors modifying this relation are listed  in
\cite{cabrera2006adventitia} where it was shown that arterial wall
viscosity and elasticity were influenced by adventitia removal in
in-vivo studies, possibly by a smooth muscle-dependent mechanism.

\subsection{Non linear wall model and evaluation of the parameters}

In a linear approach for an isotropic, incompressible and homogeneous arterial wall with a thickness $h$ and radius $R$ the thin cylinder theory stands that the 
 the stress $\sigma $ and the linearized strain $\epsilon = { R - R_0 \over R_0}$ follow the equation
\begin{eqnarray}
 \sigma={E \over (1 - \eta^2)} \epsilon
 \label{eq:1}
\end{eqnarray}
where $E$ is the Young's modulus, $R_0$ the unstressed radius and 
$\eta$ is the Poisson ratio (0.5 for incompressible materials). Also the transmural pressure $P-P_{ext}$ is related to the stress $\sigma$ by the equation 
\begin{equation} 
\sigma=\frac{R(P-P_{ext})}{h}. \label{stress}
\end{equation}  
The reader can refer to \cite{fung1997biomechanics} for details. 
Conform to the experimental setup we set the external pressure to zero and we have
\begin{eqnarray}
(1 - \eta^2) { R \over h} P = E \epsilon
\end{eqnarray}
the equation linking the pressure $P$ to the strain $\epsilon$. Adding to the right hand side a viscoelastic term $\phi \dot{\epsilon}$ where $\phi$ is the coefficient modeling the wall viscosity we retrieve a classic Kelvin-Voigt model. We can build a more general nonlinear wall model by developing $\epsilon$ and $\dot{\epsilon}$ asymptotically to  second order, we then find 
\begin{eqnarray}
(1 - \eta^2) { R \over h} P = E \epsilon + E_{NL} \epsilon^2 + \phi_0 \dot{\epsilon} + \phi_{NL} \dot{\epsilon}^2
\label{eq:general}
\end{eqnarray}
where the subscript $NL$ stand for nonlinear and where we have $ E_{NL} << E$.

We will show in the Results section that the nonlinear term in $\epsilon^2$ does not play an important role in the pressure dynamics. In fact the experimental data are in regions of small $\epsilon$ therefore $\epsilon^2 << \epsilon$ and the therm $E_{NL} \epsilon^2$ is negligible. 
  
Re-arranging the equation (\ref{eq:general}) and recalling that $E_{NL} = 0$  we  get the following relationship connecting the pressure $P$ to the  radius $R$,
\begin{equation} P=\frac{Eh}{(1-\eta^2)R_0}-\frac{Eh}{(1-\eta
^2)}\frac{1}{R}+\frac{\phi_0 h}{(1-\eta ^2)R_0}\frac{dR}{Rdt} + {\phi_{NL} h \over (1 - \eta^2) R_0^2} {1 \over R} \left(\frac{dR}{dt}\right)^2.
\label{sheep_pressure_radius}
\end{equation} 

The pressure $P$ is a linear combination of the quantities $1/R$, $(dR)/(Rdt)$, and ${1 \over R} \left(\frac{dR}{dt}\right)^2$. Therefore we estimated the coefficients of the equation by a linear regression method. 
Written in matrix form, the problem is
$
P (t)=MC\label{Matrix form},
$
where $M$ is a $N \times 4$ matrix $[(1,...,1)^T, 1/R, dR/(Rdt),{1 \over R} \left(\frac{dR}{dt}\right)^2]$,
with $N$ the number of experimental data points, $C$ the $4 \times 1$ coefficient vector $[Eh/((1-\eta^2)R_0), -Eh/(1-\eta^2), \phi_0 h/((1-\eta^2)R_0),\phi_{NL} h / (1 - \eta^2)  ]^T$. The objective cost function is
$
J(C)=\frac{1}{N}\sqrt{\sum_{i}^{N}((P_{model})_i-P_i)^2)},
$
with $P_{model}$ the pressure predicted by the model.
We assume that the columns of the data matrix are independent in the linear space and that the errors of the measurement data are independent and identically distributed. 
According to the theory of the least square method the optimal value of $C$ is  $(M^TM)^{-1}M^TP$.

In the data matrix, we evaluated the derivative of $R$ by a spectral numerical method.
Given a times series $R(t)$ with a period $T$ which is expanded in Fourier series 
$
R=\sum_{k=-\infty}^{\infty}\hat{R}(k)e^{\frac{2\pi i}{T}kt},
$
where $\hat{R}$ is 
$
\hat{R}(k)=\frac{1}{T}\int_{t=0}^{T}R(t)e^{-\frac{2\pi i}{T}kt}dt.
$
For the derivative, one has
\begin{eqnarray*}
\frac{dR}{dt}=\sum_{k=-\infty}^{\infty}\hat{R}\frac{2\pi i}{T}ke^{\frac{2\pi i}{T}kt}.
\end{eqnarray*}

In the computation, we take advantage of the Discrete Fourier Transform (DFT).
The experimental measurements are filtered out through a loop in the calculation.
The pseudo-code is:
\begin{itemize}
\item  Step 1: Evaluate the DFT of $R$ (assume $N$ as an even number without loss of generality)
$
\hat{R}_k=\frac{1}{N}\sum_{n=-\frac{N}{2}+1}^{\frac{N}{2}}R_n\cdot e^{-\frac{2\pi i}{N}nk},\quad k=-\frac{N}{2}+1...\frac{N}{2}.
$

\item Step 2: $||\hat{R}_k||$ represents the amplitude of the k-th wave.
To filter out the high frequency experimental noise, we impose a criterion $\gamma$ such that if $||\hat{R}_k||< \gamma$,  $\hat{R}_k$ is set to 0. The value of $\gamma $ will be optimized by minimizing the cost function through the loop.  

\item Step 3: Multiply  $\hat{R}_k$ by  $\frac{2\pi k i}{T}$ to get $\widehat{DR}_k$

\item Step 4: Evaluate the inverse DFT of $\widehat{DR}$
$
\left(\frac{dR}{dt}\right)_k=\sum_{k=-\frac{N}{2}+1}^{\frac{N}{2}}\widehat{DR}e^{\frac{2\pi i}{N}nk}
$

\item Step 5: Solve the least square problem and evaluate the objective function, $J(C)$.

\item Step 6: Change $\gamma$ and return back to Step 2 until the value of the objective function stops decreasing.

\end{itemize}
The thickness $h$ and the unstressed radius $R_0$ are directly measured.
From the optimization process we computed  the Young's modulus $E$ and
the viscosity coefficients $\phi$ and $\phi_{NL}$ as well as the
unsteady radius $R_0$ again. We will show the measured and optimized
values of $R_0$ are equivalent. 

We note we have performed the optimization process for a given imposed
physiological frequency given by the Jarvik device, but we know that
the model parameters depend on the frequency. Therefore
we are supposing that the model (both linear or nonlinear) is valid
for all frequencies, and this is a strong hypothesis when we use only
one imposed frequency. To confirm that hypothesis we will need to design
an optimization process for large band frequencies and show the
optimal parameters are independent of the input frequency.
Finally, once the optimal coefficients found we can introduce them
into the numerical model to study the wave propagation in the network.

\subsection{Simulation of pulse waves with the 1D model}

For blood flow in arteries, denoting the circular cross-sectional area by $A$, the flow rate by $Q$ and the internal pressure by $P$, the conservation of mass and balance of momentum follow two partial differential equations (PDEs):
\begin{eqnarray}
\frac{\partial A}{\partial t}+\frac{\partial Q}{\partial x}=0, \label{massConserv_AQ}\\
\frac{\partial Q}{\partial t}+ \frac{\partial}{\partial x} \left(\frac{Q^2}{A}\right)+ \frac{A}{\rho}\frac{\partial P}{\partial x} = -C_f \frac{Q}{A}, \label{momentumConserv_AQ}
\end{eqnarray}
where $x$ is the axial distance and $t$ is the time.  
The blood density $\rho$ is here constant, and $C_f$ is the skin friction coefficient which depends on the shape of the velocity profile. In general, the profile depends on the Womersley number, $R\sqrt{\omega/ \nu}$, with $\omega$ the angular frequency of the pulse
wave and $\nu$ the kinematic viscosity of the fluid.
In practice, $C_f$ usually takes an empirical value fitted from experimental observations.
In this study, we assume $C_f=22\pi\nu$ as fitted for the blood flow in large vessels with a Womersley number of about 10~\cite{smith2002anatomically}. 
To close the system of equation we need a  constitutive equation for the pressure $P$, the equation (\ref{sheep_pressure_radius}), is then written 
\begin{equation} 
P=P_{ext}+\beta(\sqrt{A}-\sqrt{A_0})+\nu_s \frac{\partial A}{\partial t} + \nu_{NL} \left(\frac{\partial A}{\partial t}\right)^2,
\label{sheep_constitutive}
\end{equation} 
with
\[ \beta=\frac{\sqrt{\pi} Eh}{(1-\eta^2)A_0}, 
\quad 
\nu_s=\frac{\sqrt{\pi}\phi_0 h}{2(1-\eta^2) A \sqrt{A_0}}
\quad
\nu_{NL}  ={1 \over 4} { \phi_{NL} h \over ( 1 - \eta^2) } { \pi^{1/2} \over A^{3/2} A_0} 
\] 
the elastic and viscoelastic coefficients. Those are the governing
equations for the blood flow in one segment.

In the numerical simulations we preserved the conservation of mass and static
pressure at the confluence points between segments. We neglected the
energy loss due to the variation of geometry. We impose a flow rate as input condition at
the inlet of the network (Ascending Aorta). The flow rate is a cyclic
half sinusoidal function in time of a period of 0.5~s.; the simples
one having the right signal characteristic in terms of physiological amplitude and
frequency. A more physiologic waveform shape does not change significantly
the numerical results, which are primarily affected by the confluence
reflexions and terminal resistances.
At each outlet, we imposed an identical small reflection coefficient $R_t = 0.3$.

We solved the governing equations numerically using a finite volume approach by a Monotonic Upstream Scheme for Conservation Laws (MUSCL).
The code has been favorably validated with analytic results and experimental data, see~\cite{wang2012comparing,wang2014verification}.

\section{Results and discussion}

\label{sec:results-discussion}

\subsection{Parameters of the arterial wall}

Before the complete presentation of the nonlinear Kelvin-Voigt model optimization results we discuss the differences in predictions when using a classical linear model and the relative importance of the nonlinear term in $\epsilon^2$.

\begin{figure}[htb]
	\includegraphics[width=0.5\textwidth]{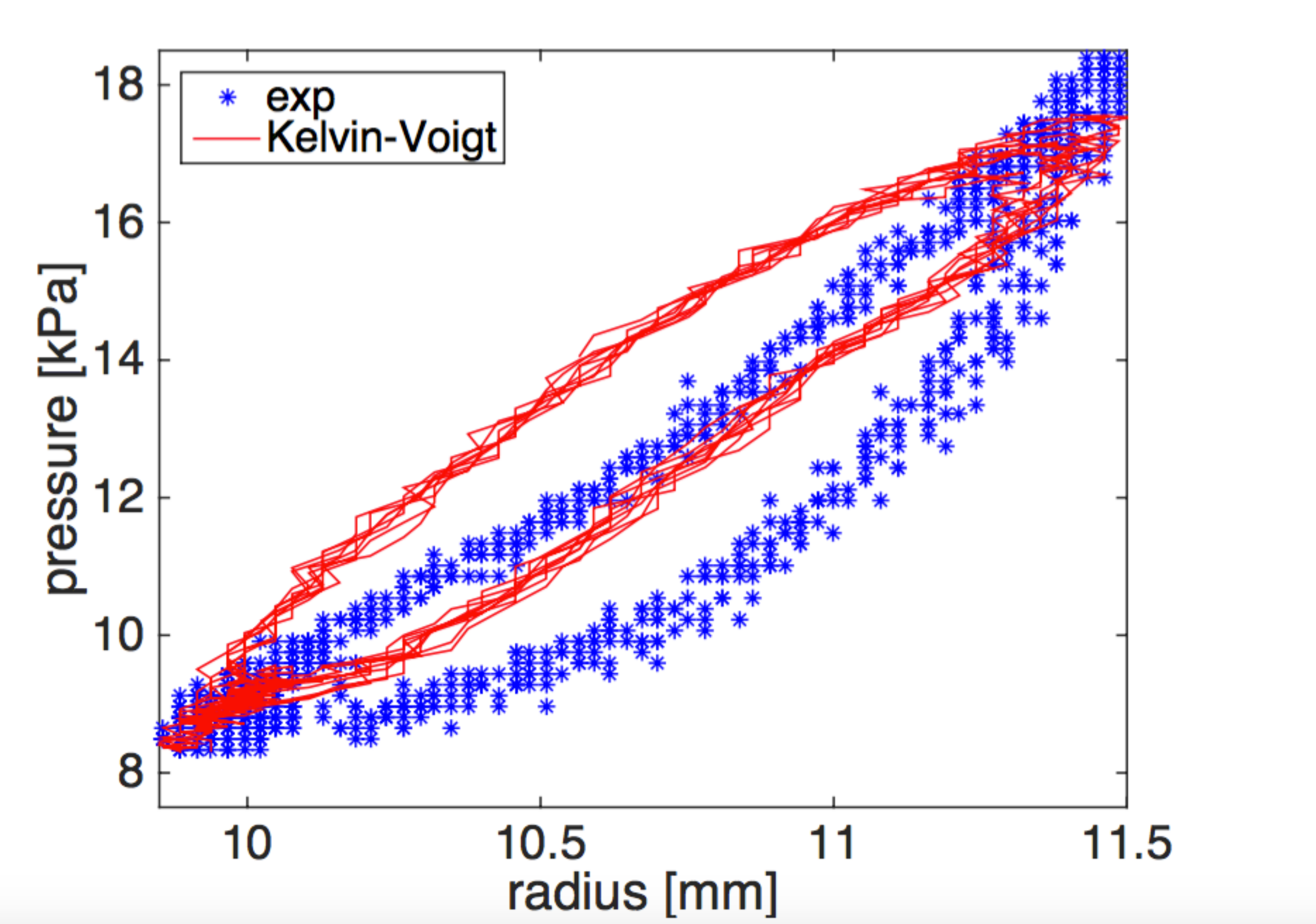}
	\includegraphics[width=0.5 \textwidth]{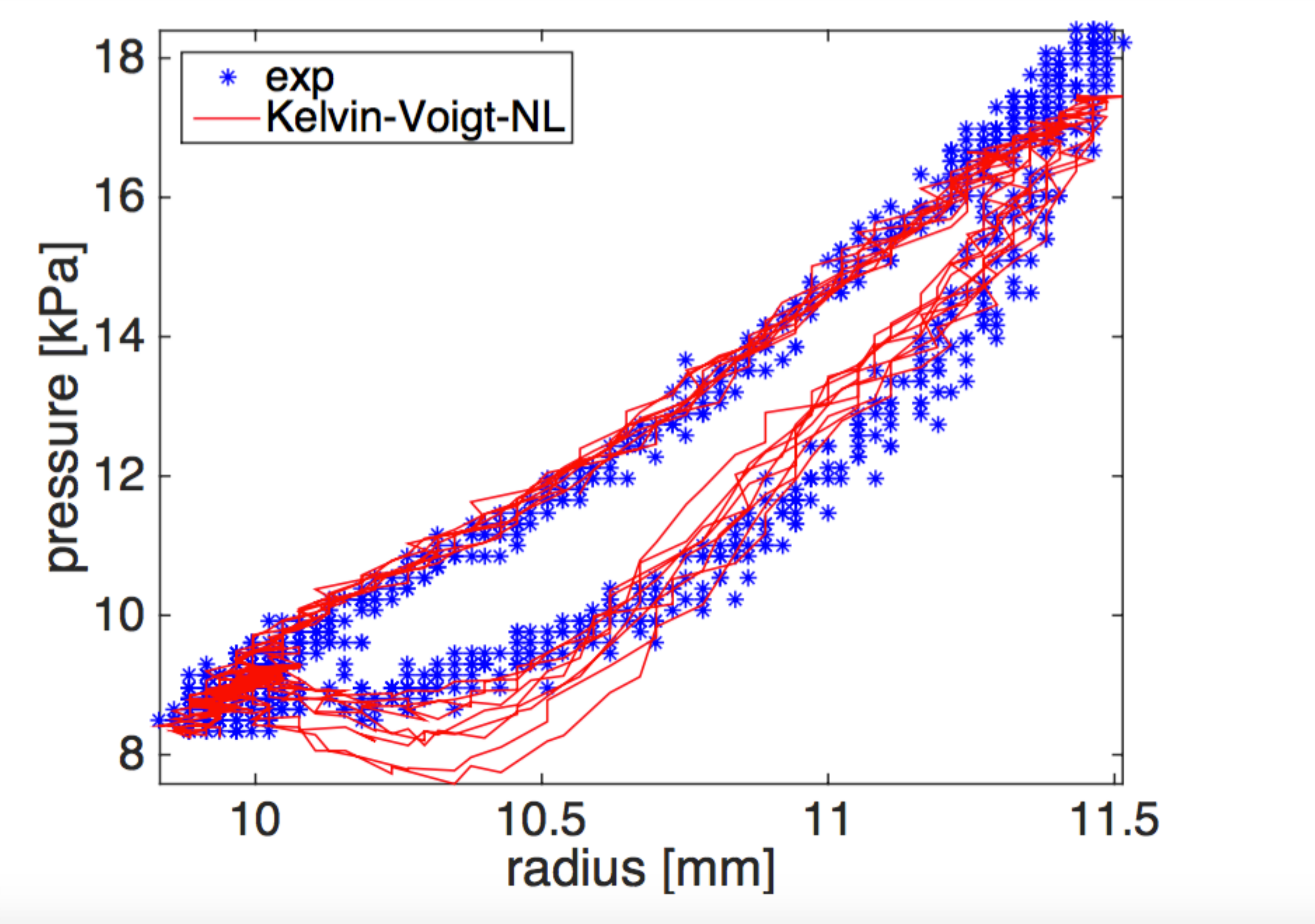}
	\caption{Pressure-radius loop of Ascending Aorta : Experimental data and prediction of (left) linear Kelvin-Voigt model and (right) nonlinear  Kelvin-Voigt model.}
	\label{fig:comparison}
\end{figure}

The Figure \ref{fig:comparison} presents in the Ascending Aorta the prediction of the linear model ($E_{NL}$ and $\phi_{NL}$ set to zero) and the nonlinear Kelvin-Voigt model proposed 
\begin{eqnarray}
(1 - \eta^2) { R \over h} P = E \epsilon + \phi_0 \dot{\epsilon} + \phi_{NL} \dot{\epsilon}^2.
\label{eq:model}
\end{eqnarray}
The linear model fits poorly the curvature observed in the experimental data (Figure \ref{fig:comparison} (left)), and is equivalent to the 
Valdez-Jasso et al.~\cite{valdez2009analysis} optimization  analysis that adopts a stress relaxation constant as an extra parameter into a linear Kelvin-Voigt model. On the contrary the nonlinear prediction (Figure \ref{fig:comparison} (right)) properly follows  the experimental data. We note that linear and nonlinear optimal parameters for ($E,\phi_0$) computed independently are very similar : linear ($1.475 \ MPa, 26.156 \ KPa \cdot s$) and nonlinear ($1.539 \ MPa, 25.451 \ KPa \cdot s$).

From the experimental data of Figure \ref{fig:comparison} we can
evaluate the order of magnitude of $\epsilon$ which is around
$1.5/10$, therefore the nonlinear term scales as $\epsilon^2 \sim 2 \
10^{-2}$. As the ratio $E_{NL} \over E$ is around $10^{-2}$ as shown
by the numerical results gathered from the optimization process with
the nonlinear parameter $E_{NL}$, the linear term scales as $E
\epsilon$ and the nonlinear one as $10^{-3} E \epsilon$. The
numerical predictions using the nonlinear term $E_{NL}$ confirm that
this nonlinear term as small influence and can be neglected as already advanced in \cite{segers1997assessment}.

Over the subsequent optimizations we used the nonlinear Kelvin-Voigt model (equation \ref{eq:model}). As stated in the introduction  there are other models for the viscoelasticity (see e.g.~\cite{bessems2008experimental,holenstein1980viscoelastic,segers1997assessment,valdez2011linear}).
Fung's quasilinear model is more generalized than the spring-dashpot models, but its incorporation in 
1D fluid models is complex, thus it is only applicable to limited formulations (e.g. linearized 1D model~\cite{holenstein1980viscoelastic,segers1997assessment}).

  \begin{figure}[htb]
 \includegraphics[width=0.5\textwidth]{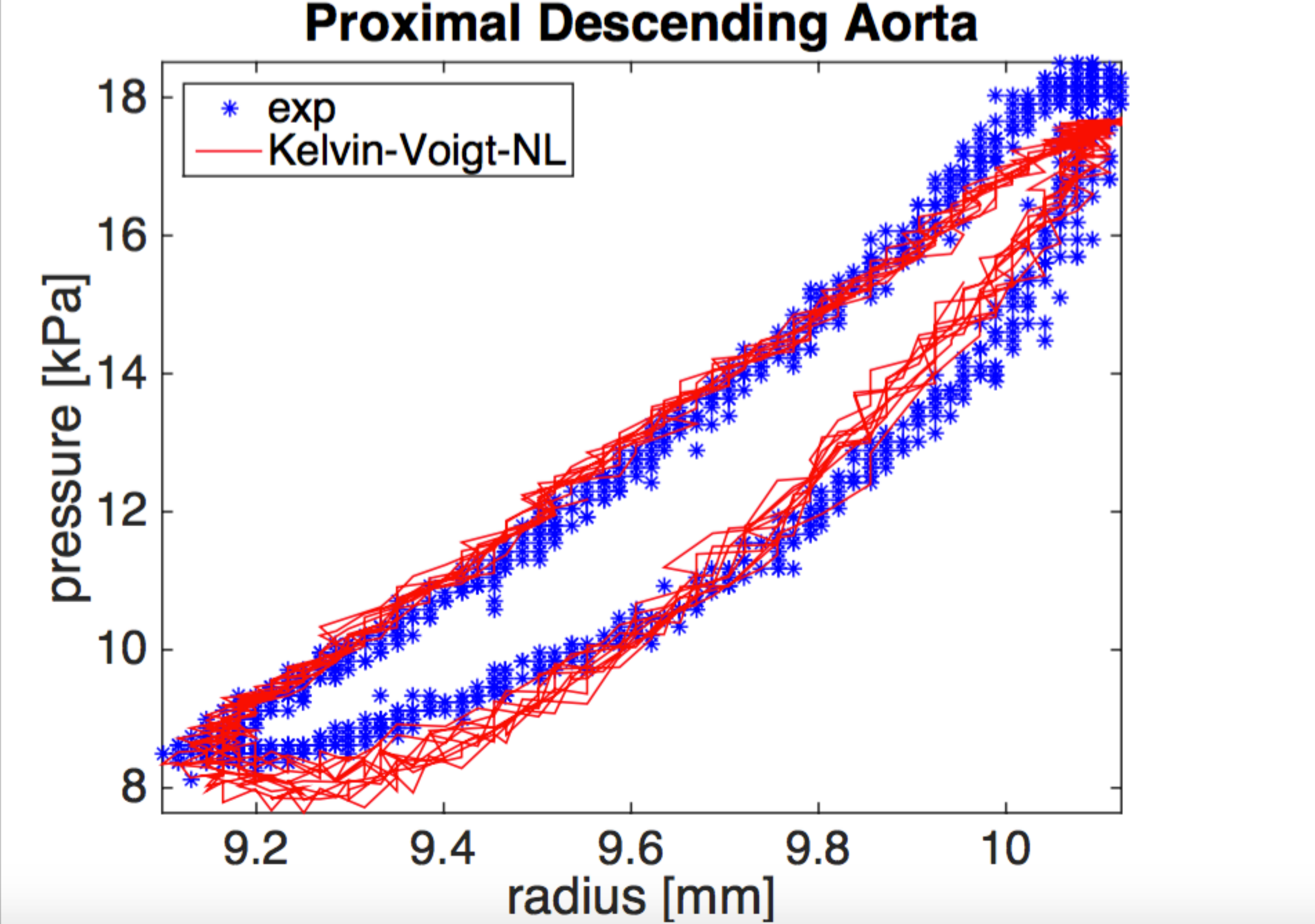} \includegraphics[width=0.5 \textwidth]{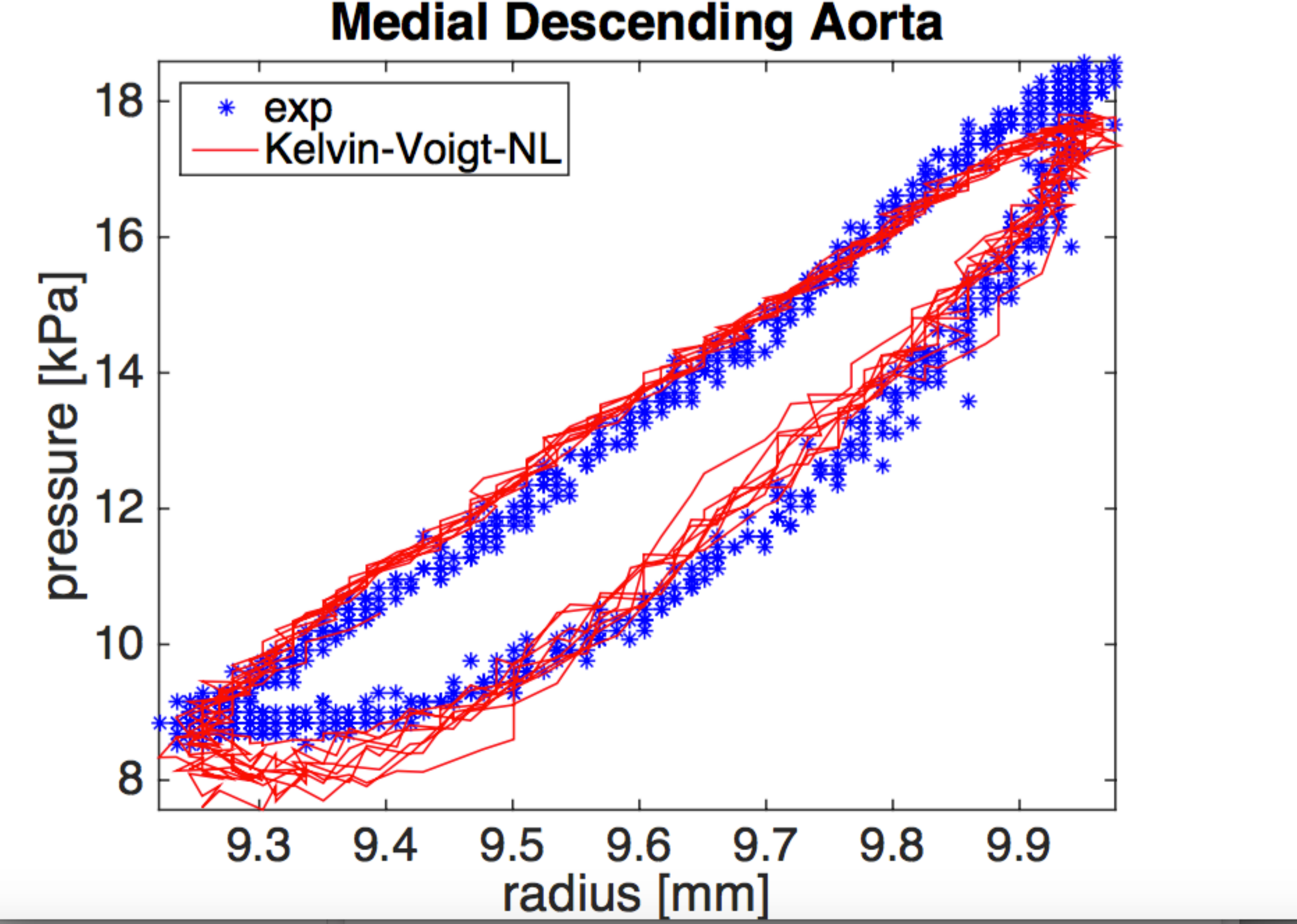}

  \includegraphics[width=0.5\textwidth]{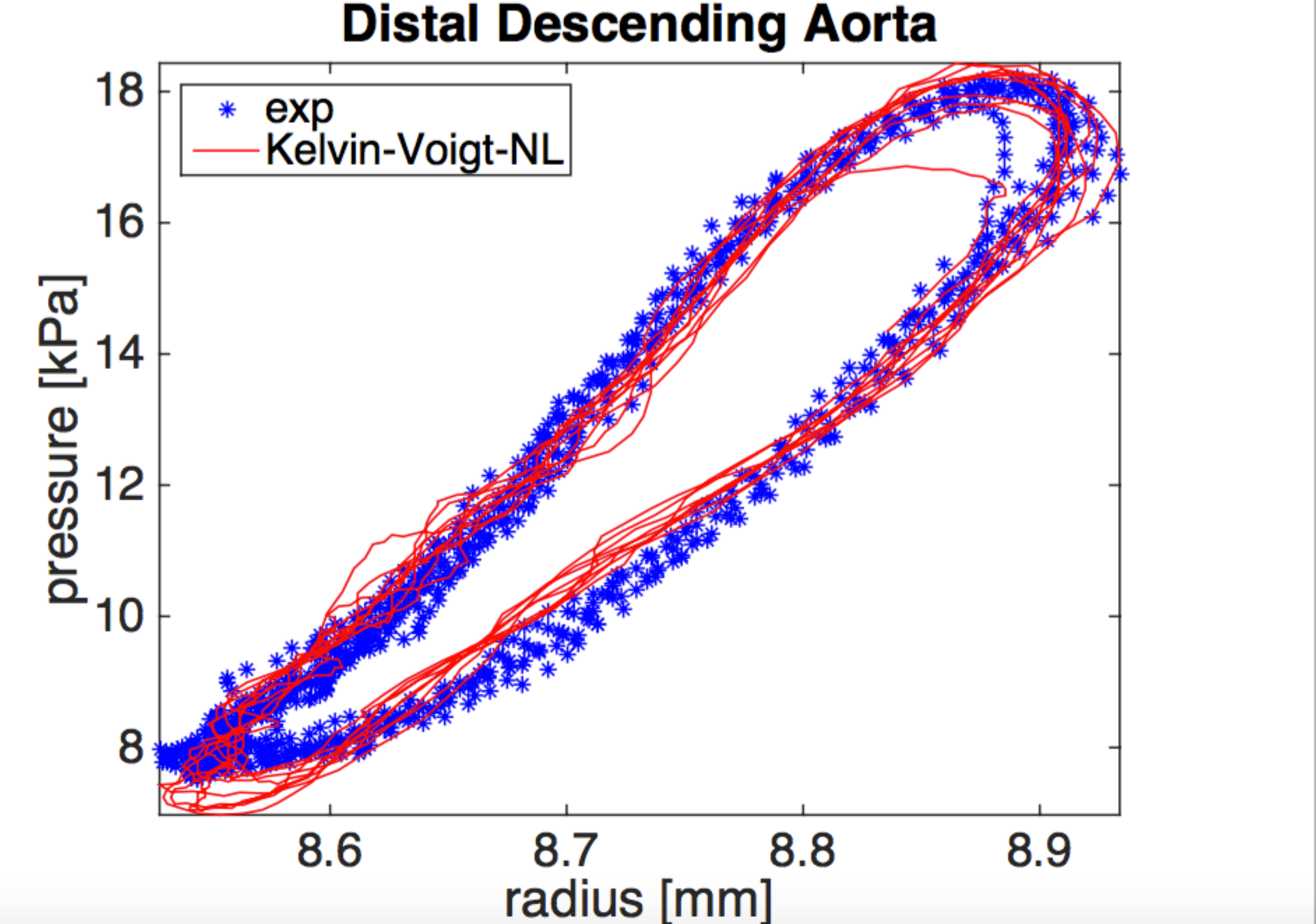} \includegraphics[width=0.5\textwidth]{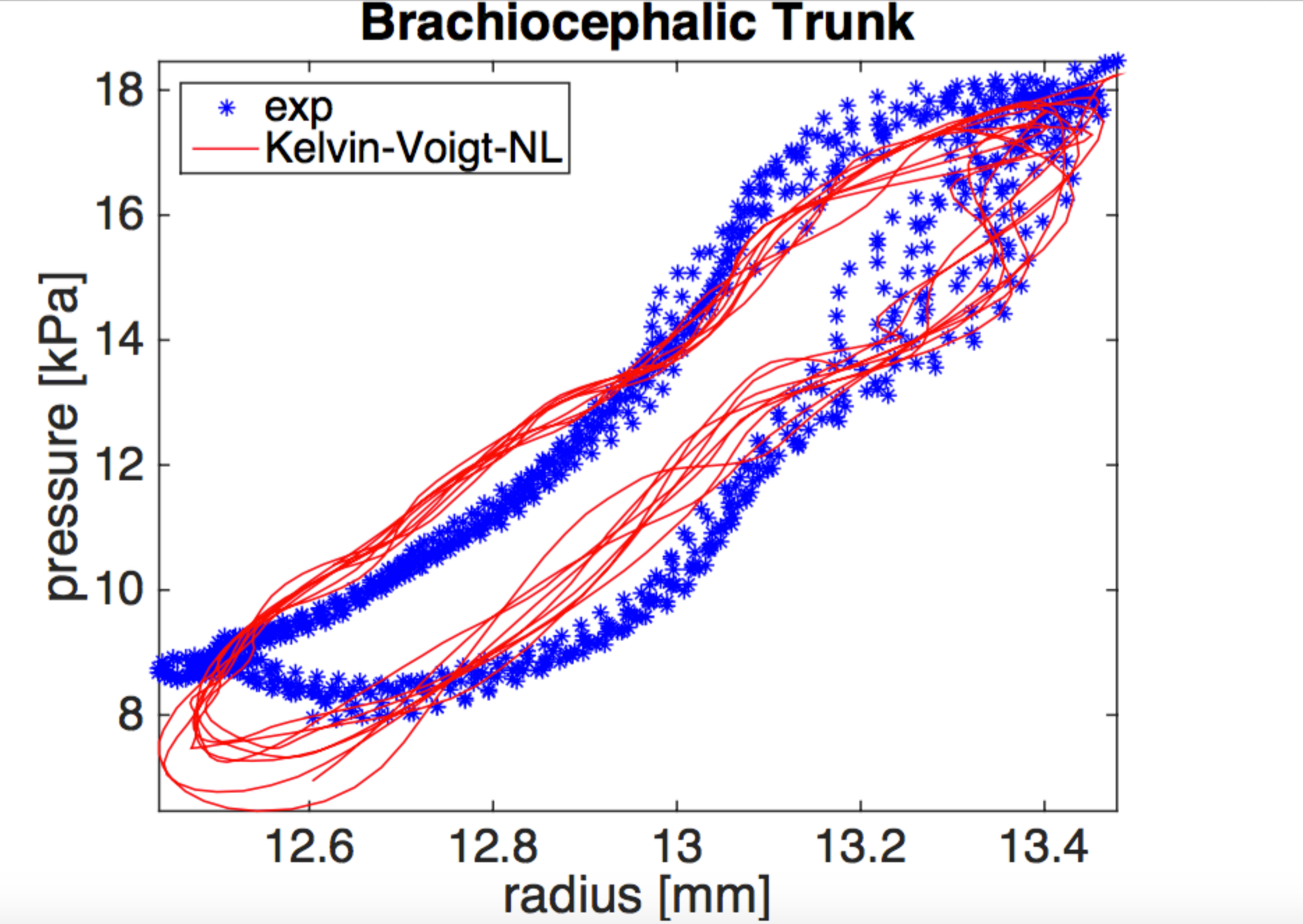}
  			   
 \includegraphics[width=0.5\textwidth]{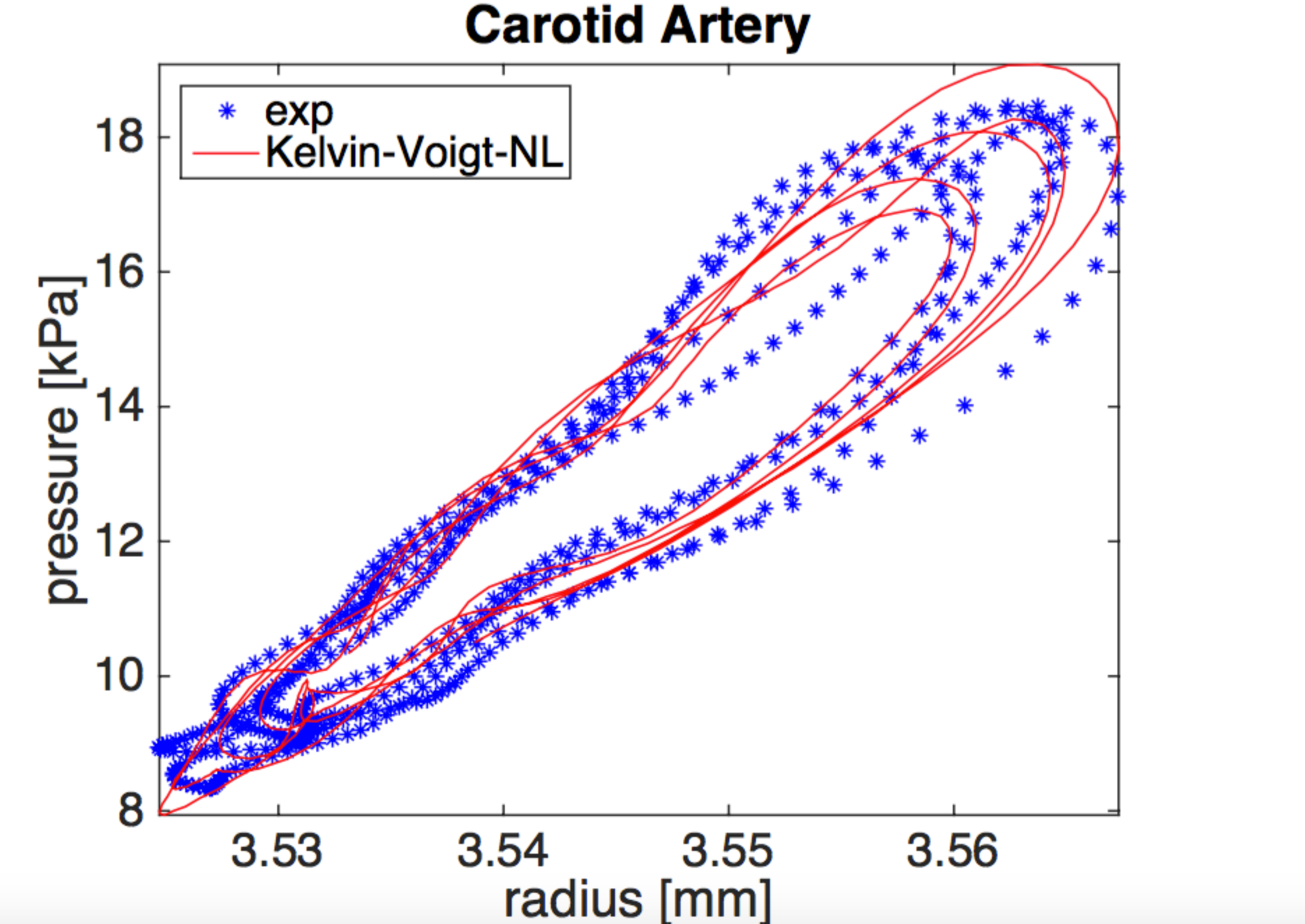}
 \includegraphics[width=0.5 \textwidth]{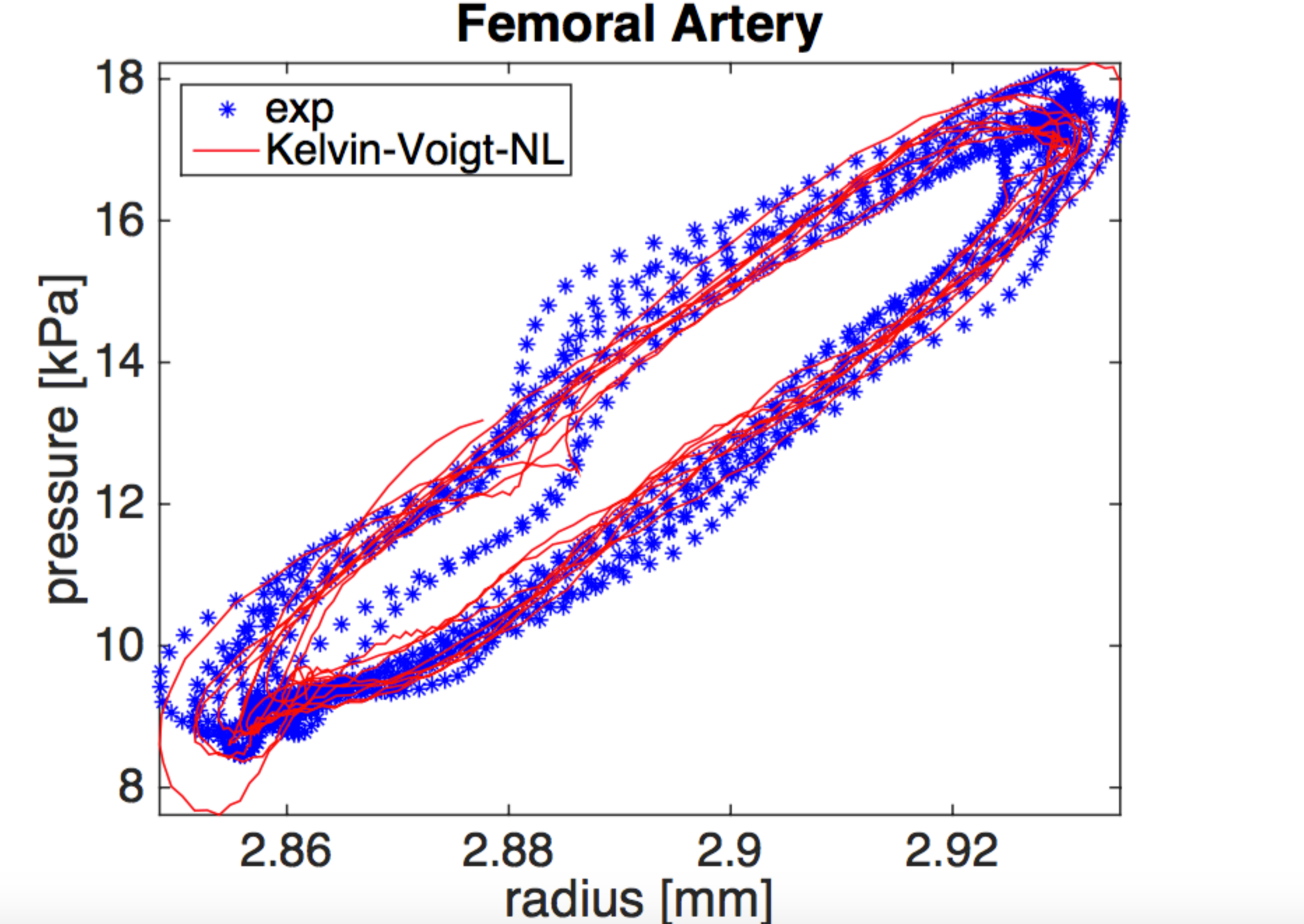}
  				\caption{Experimental data and the fitted nonlinear Kelvin-Voigt model. Parameter values are in Table\ref{tab:Parameters_sheep_tree}.}
  				\label{fig:results}
  \end{figure}

The results presented in Figure \ref{fig:results} from the upper left side to the bottom right side (Proximal Descending Aorta, Medial Descending Aorta, Distal Descending Aorta, Brachiocephalic Trunk, Carotid Artery, and Femoral Artery) show that this model captures the wall viscosity and the nonlinearity of the pressure-radius loop.
As stated above, Valdez-Jasso et al.~\cite{valdez2009analysis} already tested the Kelvin model modeling two stress relaxation constants and their results are close to the linear Kelvin-Voigt model.
Their sensitivity analysis shows that the model prediction depends least on this constant among all the parameters, thus even though the Kelvin-Voigt does not include this constant the validity is hardly influenced.    
Moreover, in contrast to nonlinear optimization methods that estimate the model parameters in~\cite{valdez2009analysis},
we use the linear regression method which is fast and the global optimization is readily guaranteed.

Figure~\ref{fig:results} shows the hysteresis in the pressure-radius loop for six arteries (the seventh, the Ascending Aorta is in Figure \ref{fig:comparison}).
The agreement between the experimental measurements and the model predictions shows that the nonlinear Kelvin-Voigt model captures the wall viscosity everywhere. 
We remark that among the seven arteries, the brachiocephalic trunk has the largest nonlinearity (Figure~\ref{fig:results} center and left).
	
At the aorta, the nonlinearity decreases from the proximal part to the distal end. Finally at the peripheral arteries, represented by carotid artery and femoral artery, the nonlinearity is negligible.

\begin{figure}[h]
	\centering
\includegraphics[width=0.48\textwidth]{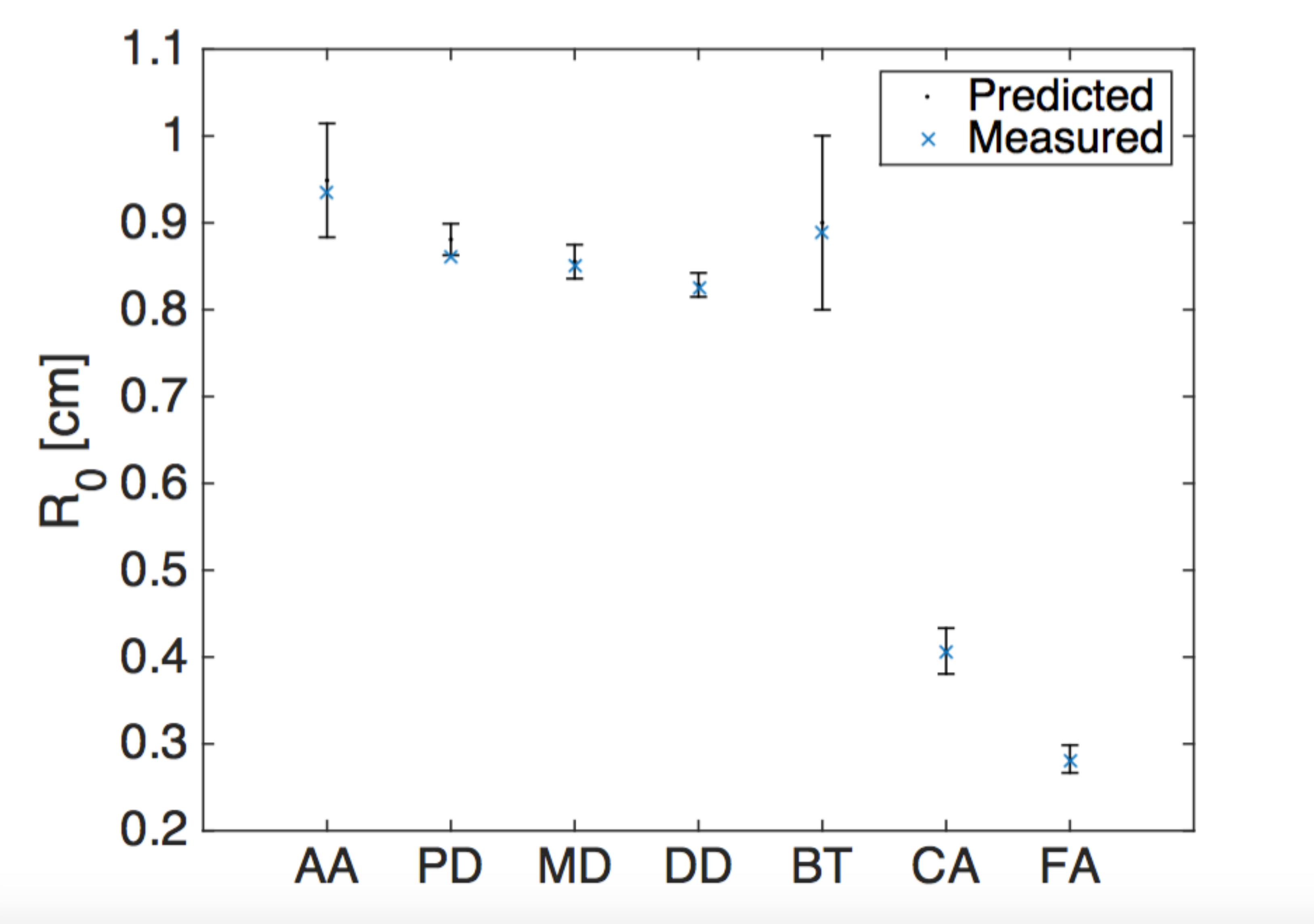} 

\caption{Optimal unstressed radius $R_0$, predicted and measured for the seven arteries (Ascending Aorta, Proximal Descending Aorta, Medial Descending Aorta, Distal Descending Aorta, 
	Brachiocephalic Trunk, Carotid Artery, Femoral Artery)}
\label{fig:r0}
\end{figure}

We present the unstressed ratio $R_0$ with error bars  in Figure \ref{fig:r0} and the mean values are [Ascending Aorta, Proximal Descending Aorta, Medial Descending Aorta, Distal Descending Aorta, Brachiocephalic Trunk, Carotid Artery, Femoral Artery] = $[0.9489, 0.8809, 0.8554, 0.8286, 0.9002, 0.4069,0.2826]$. These values
 compare extremely well the experimentally measured ones $[0.9360,
 0.8600, 0.8500, 0.8250, 0.8900,$ $0.4060, 0.2810]$ (crosses in
 Figure). The experimental measurements of neutral vessel radius are
 only possible in \textit{in vitro} experiments but impossible in an
 \textit{in vivo} analysis, this is the reason we chose to estimate
 the values of the radius $R_0$ in numerical simulations. Since the
 nonlinear Kelvin-Voigt model predicts the actual values (within the
 error bars), that suggests that this approach could be used in an
 \textit{in vivo} situation.

\begin{figure}[htb]
	\includegraphics[width=0.48\textwidth]{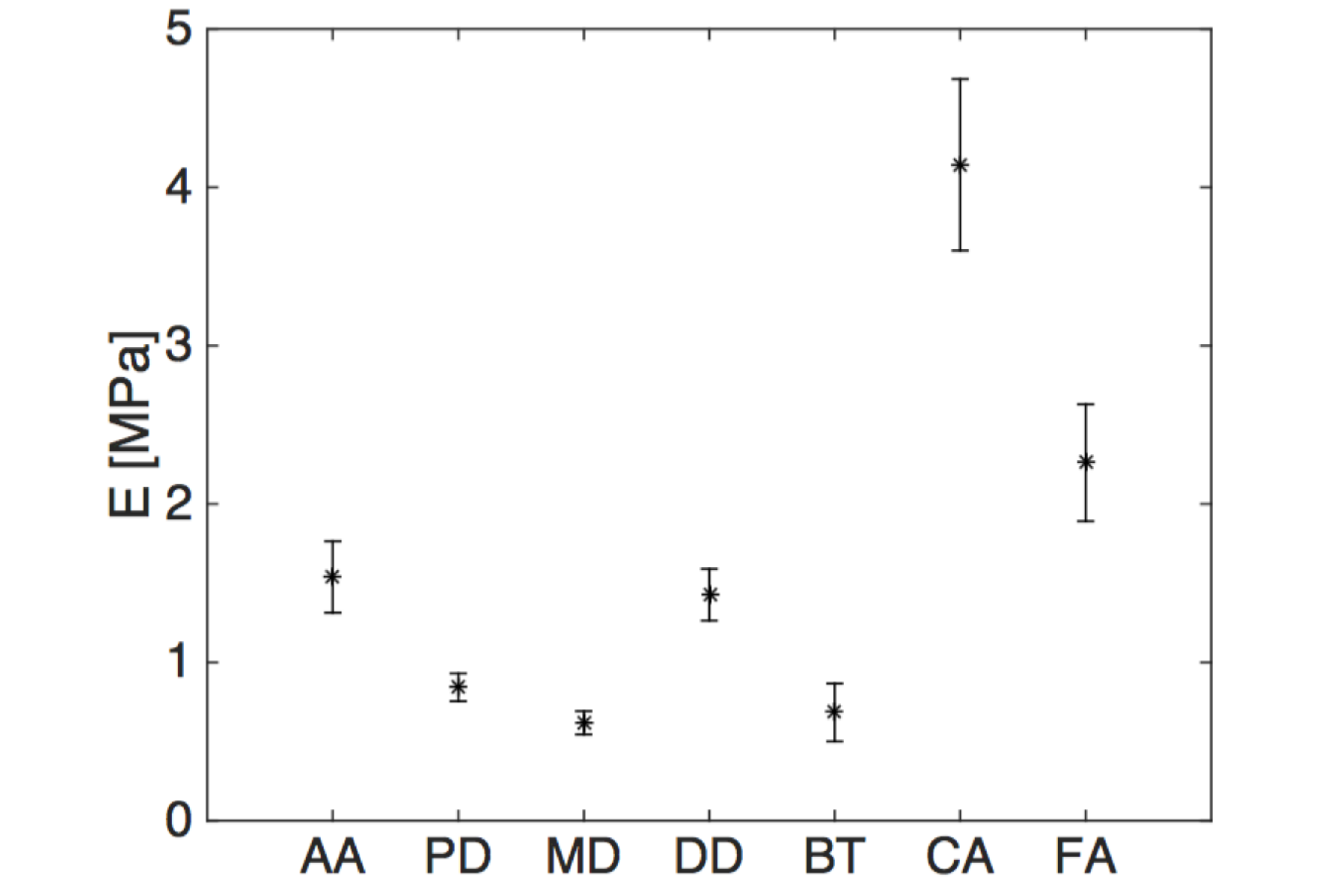} 
	\includegraphics[width=0.48\textwidth]{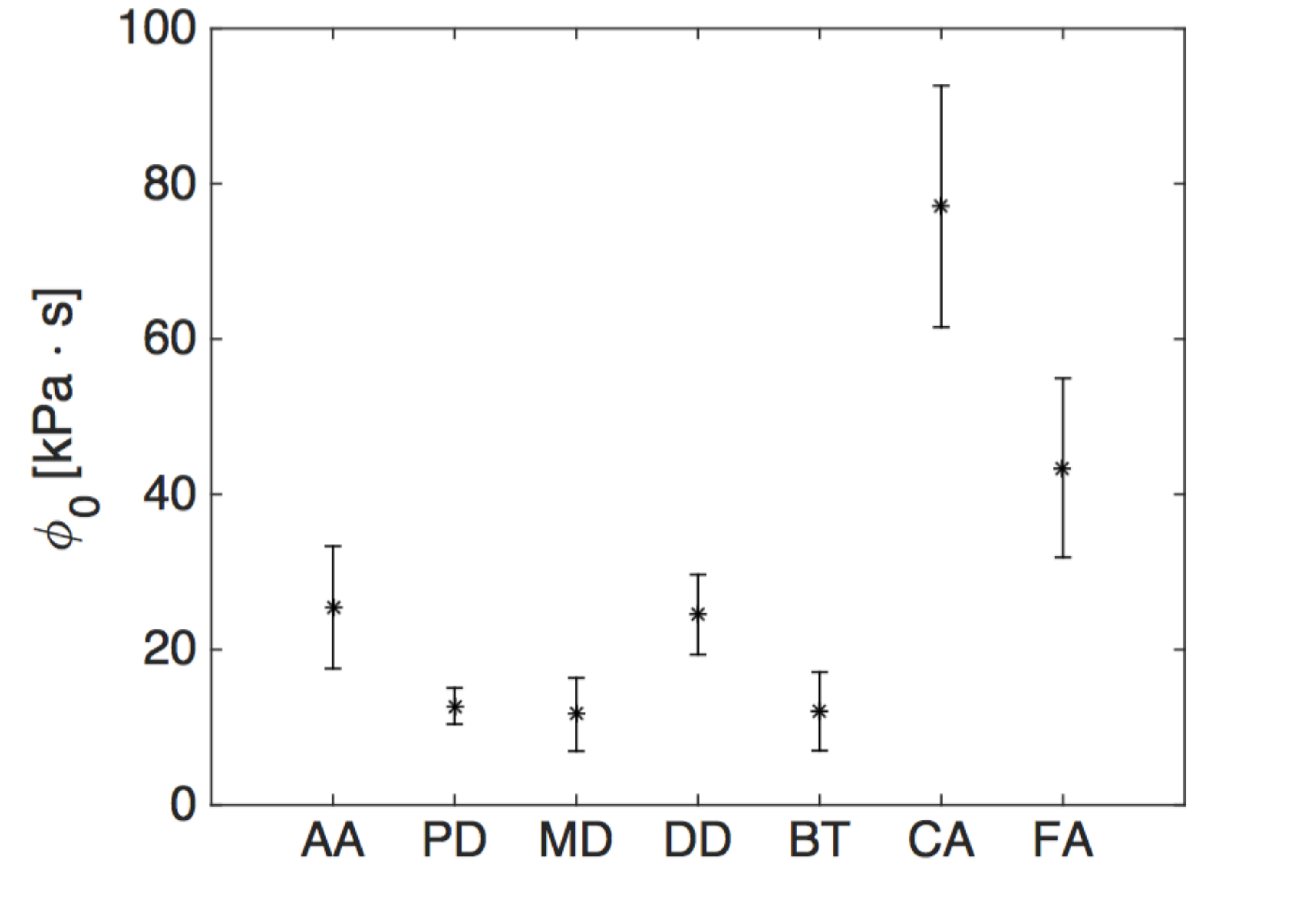}   

	\caption{Mean values of the reference Young's modulus $E$ (left), and 
		viscosity coefficient $\phi_0$ (right) with standard deviations among the group of sheep at the seven locations of the arterial network.}
	\label{fig:visco}
\end{figure}

Figure \ref{fig:visco} (left) shows the Young's modulus for the seven arteries and these results have to be compared to those of the linear viscoelastic modulus $\phi$ in Figure \ref{fig:visco} (right). Both predicted values follow the same behavior. 
By examining the parameter values among the different arteries, we can see that smaller arteries tend to be stiffer,
as pointed out by previous
studies~\cite{valdez2011linear,valdez2009analysis}. We note that
running the optimal process for the linear model we found similar
values of $E$ and $\phi_0$, this implies that these parameters are
unaffected by the nonlinear coefficient and suggests that at the
first order they have a physical meaning. 

\begin{figure}[htb]
	\centering
	\includegraphics[width=0.48\textwidth]{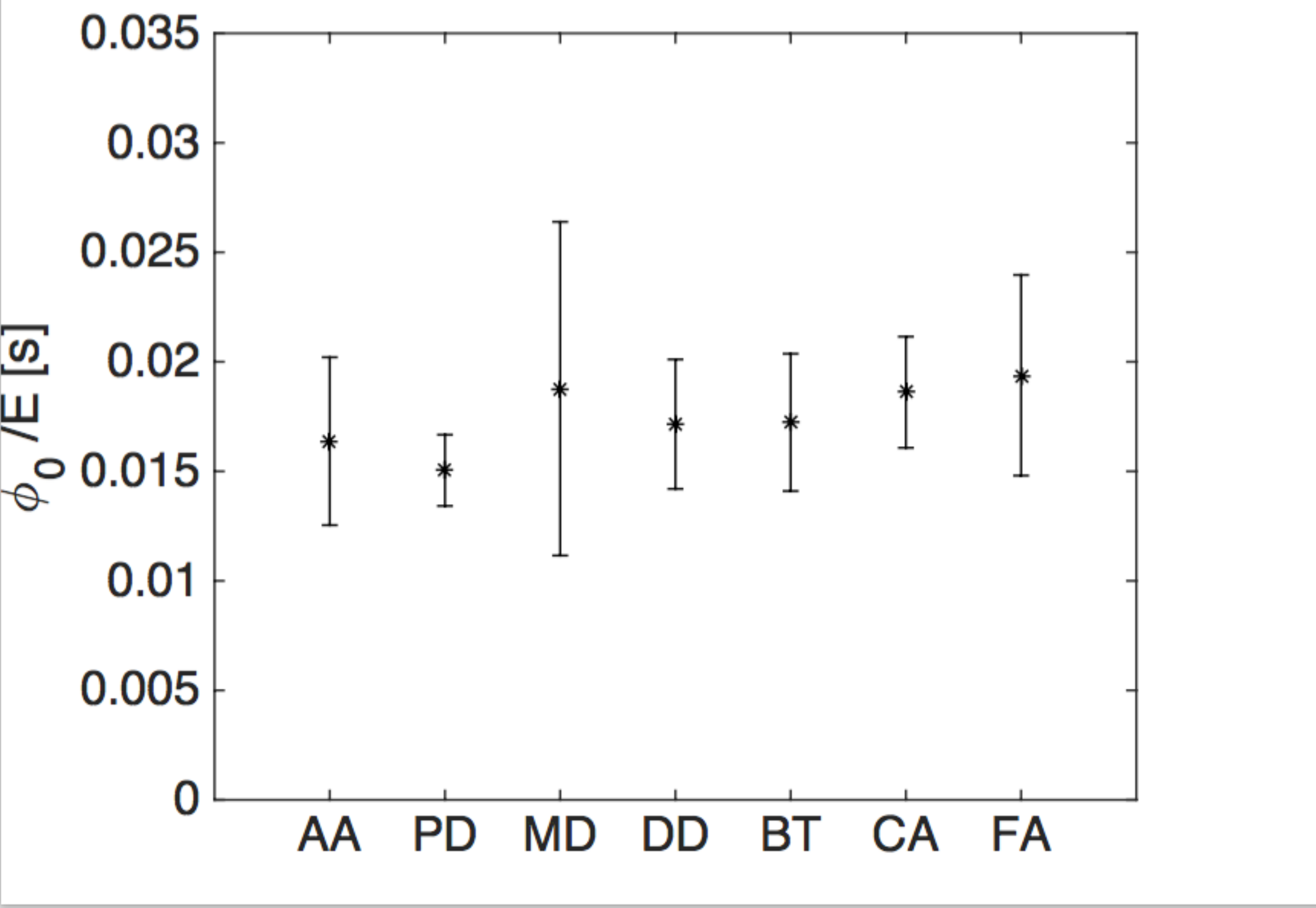} 
	\caption{Relaxation time $\phi_0/E$  with standard deviations at the seven locations of the arterial network.}
	\label{fig:ratio}
\end{figure}

We analyzed the relation between the Young's modulus and the
viscoelastic coefficient by defining a characteristic time ${\phi_0
  \over E}$. The Figure \ref{fig:ratio}  presents the values for the
seven arteries: an important observation is that these values seem to be constant. 

We want to stress in this study the importance that the ratio between
$\phi_0$ and $E$ is constant and the impact this has on high
frequencies components of the pulse waves. From the linear
Kelvin-Voigt equation~(\ref{eq:model}), with $\phi_{NL}=0$, the
magnitude of the complex modulus is
$ |G| = E \sqrt{1 + \left(\frac{t_r}{t_f}\right)^2}$, where
$t_r = \phi_0/ E $ the viscoelastic relaxation time and
$t_f = 1 / \omega$ is the typical forcing time. The linear model also
gives the phase shift as
$\delta = \arctan \left(\frac{t_r}{t_f}\right)$. Therefore for a
imposed pressure perturbation on a viscoelastic arterial wall, the
wall will come back to its equilibrium state but with a phase lag
characterized by the viscoelastic relaxation time.
Figure~\ref{fig:ratio} shows that the viscoelastic relaxation time
seems to be a biological constant. It is then evident that for high
values of $\omega$ the pressure perturbations vanish as long as
$\arctan \left(\frac{t_r}{t_f}\right) \rightarrow \pi /2 $. The
pressure perturbation and the wall response will be in phase
opposition. This indicates that higher frequency of the waves will
lead to stronger damping effect of the wall viscosity. Since the
wavefronts are more steepened toward the peripheral part of the
arterial tree due to the advection effect of blood flow, the damping
effect is more significant in this part. Damping effect  is maybe a protective
factor of the micro-circulatory system.

In a stiffer vascular network, pulsatile energy at high frequency
tends to be damped in micro-circulation, especially in the brain and
kidney~\cite{nichols2011mcdonald}. The arterial wall is mainly composed of elastin and
muscular fibers and this composition varies throughout the whole
network, from the aorta to the peripheral arteries. The elastin is
more related with the elasticity modulus and the muscular fibers to
the viscoelasticity. The smaller arteries usually have more muscular
fibers than large arteries, and this may also be explained by the need
of a stronger damping factor of pulsations right before the
micro-circulations.

Finally the mean values of the ratio $ {\phi_{NL} \over \phi } $ are for Ascending Aorta, Proximal Descending Aorta, Medial Descending Aorta, Distal Descending Aorta, Brachiocephalic Trunk, Carotid Artery, and  Femoral Artery] equalt to  $[-0.915 , -0.999 , \\ -0.888,    -0.975 ,  -1.380  ,  0.524 ,   -0.395 ]$.

\subsection{Pulse waves}
\label{sec:numer-compr}

We propose a 1D numerical model to put forward the differences between
an elastic and a viscoelastic wall model. We set the nonlinear
viscoelastic coefficient $\phi_{NL}$ to zero for simplicity.
Preliminary simulations show that the behavior is similar and not
particular shape or pattern was found. The nonlinear term could play
a role in a transient state in large networks.

\begin{table}[!ht]
	\centering
	\begin{tabular}{ c c c c c c }
		& $L$ &  $R_0$   & $h$ & $E$ & $\phi_0$  \\
		Artery  & (cm) & (cm) & (mm) & (MPa) & (kPa$\cdot$s) \\  
		\hline
		AA&  4 & 0.948 &0.38 & 1.539   &  25.451 \\
		PD & 10 &  0.880 &0.91 & 0.842  & 12.746  \\
		MD & 10 &  0.855  &1.26 & 0.617  &  11.651  \\
		DD & 15 & 0.828  & 1.10 & 1.427 &   24.514 \\
		BT & 4 &   0.900 & 1.06 &  0.683  &  12.048 \\
		CA & 15 & 0.406 &0.78 &   4.142 &  77.082 \\
		FA & 10 & 0.282  &0.31 &  2.260 &   43.426  \\
		VA1 & 20 &  0.384  &  0.50  &  4.121 & 10.000  \\
		VA2 & 20 &  0.387  &  0.50  &  0.237 & 10.000  \\
		VA3 & 20 &  0.817  &  0.50  &  3.636 & 10.000     
	\end{tabular}
	\caption{Parameters of the simulated arterial tree. The length
          $L$ is from literature and the thickness $h$ is directly measured.
From the optimization process we computed  the Young's modulus $E$ and
the viscosity coefficients $\phi_0$ and the
neutral radius $R_0$. }
	\label{tab:Parameters_sheep_tree} 
\end{table}

We use the mean values coming from the optimization process. 
Table~\ref{tab:Parameters_sheep_tree} shows the parameters of the simulated arterial tree where the length $L$ of each artery is estimated from data in literature~\cite{fung1997biomechanics}.

To model the terminal branches of the aorta, we added three virtual arteries at the ends of Proximal Descending, Medial Descending and Distal Descending aorta respectively (see Figure~\ref{sheep_tree}).
We have determined the radius of the virtual arteries by  Murray's
law and we have calculated their elasticity using a well-matched condition which is essentially no reflections at the bifurcations. 

At the inlet of the network (Ascending Aorta), the flow rate is a cyclic half sinusoidal function in time with a period of 0.5~s and the peak value is $Q_{max}=55 cm^3.s^{-1}$. As long as the pressure waves travel in the network, high frequency components appear in the signal due to reflexions and the branching points and because the vessel segments are short, geometrically reducing the wavelength of the pulse waves.

\begin{figure}
\includegraphics[width=0.48\textwidth]{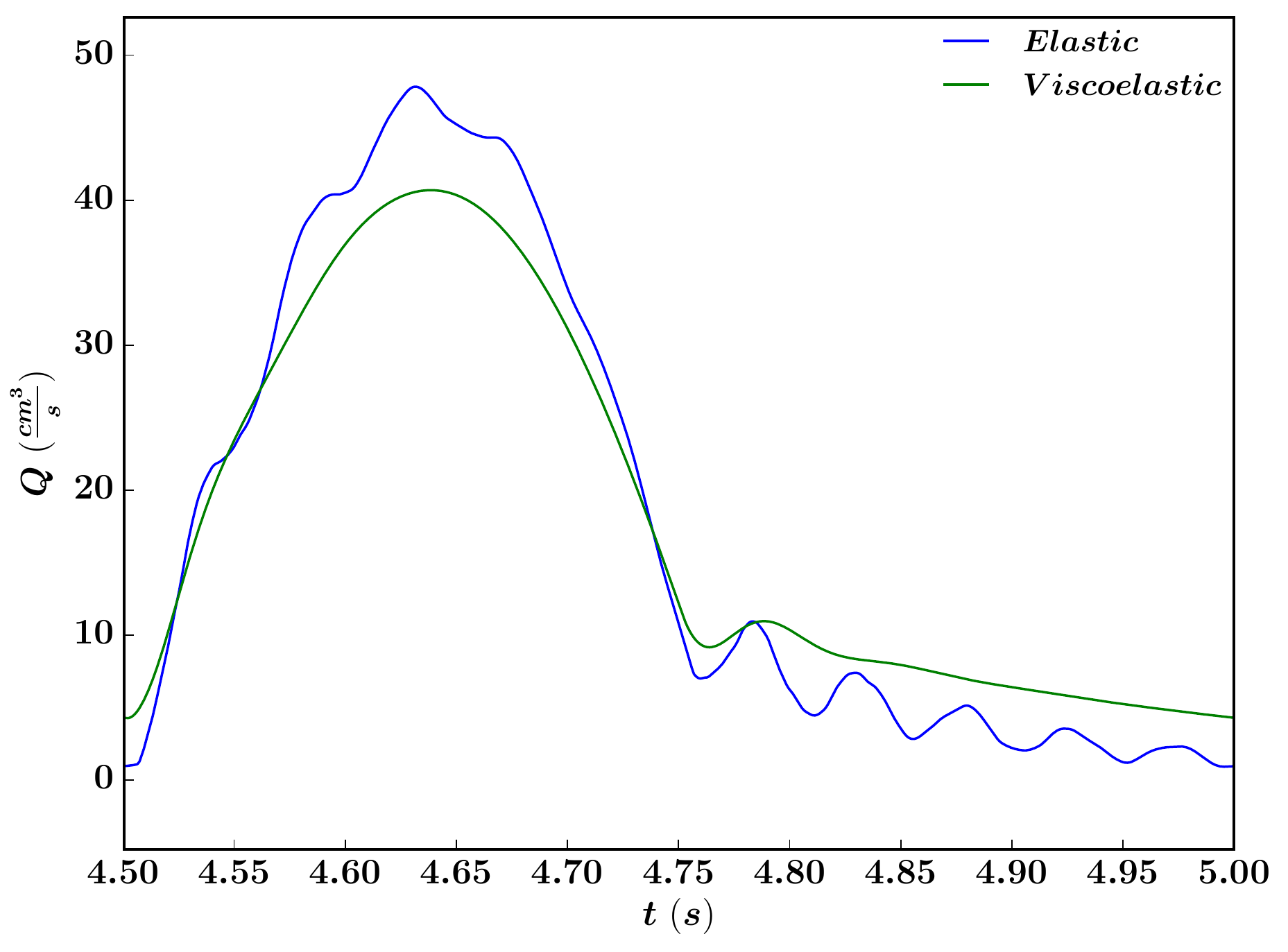}
\includegraphics[width=0.48\textwidth]{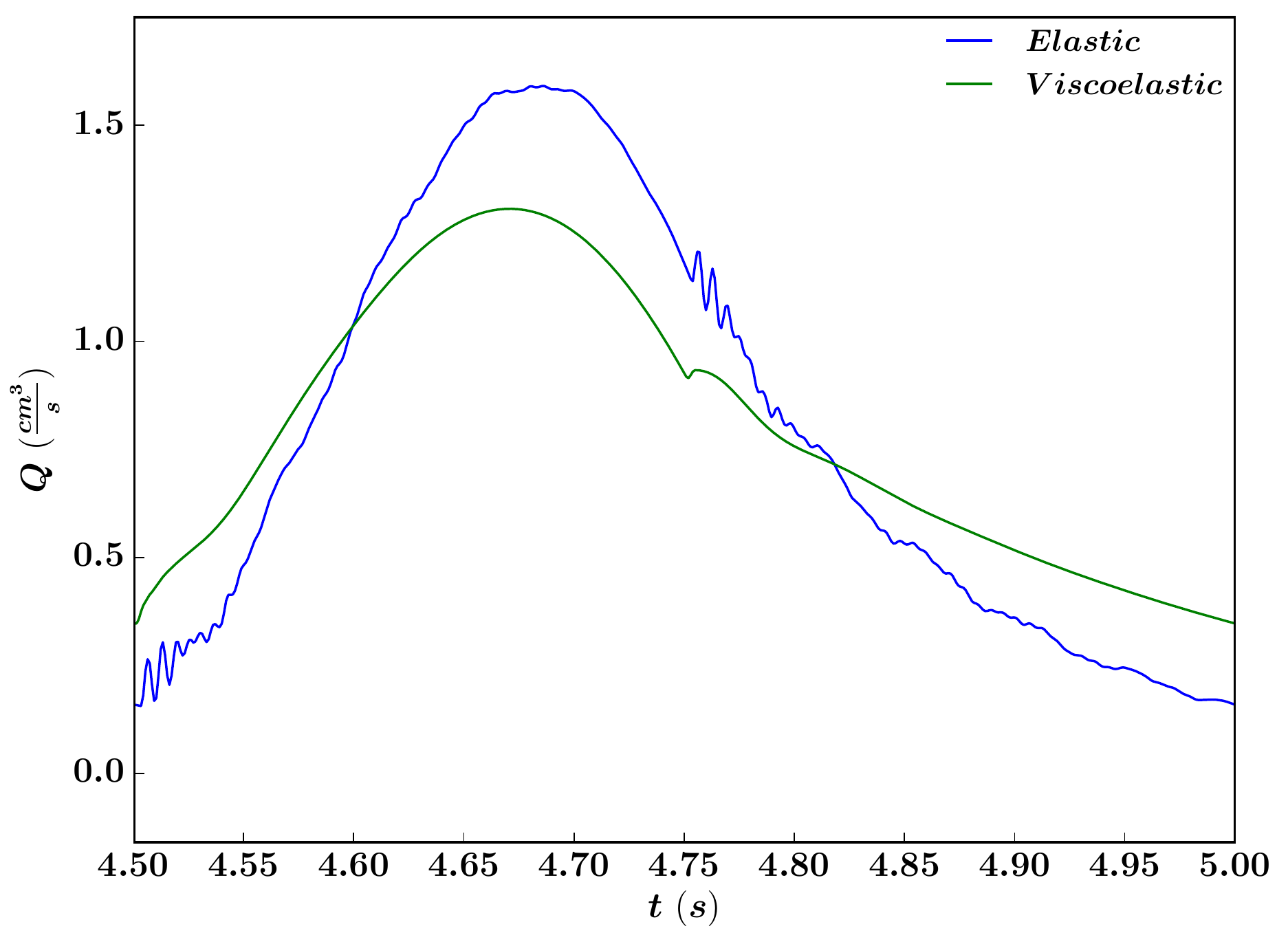}

\caption{Time series of low rate at Medial Descending Aorta (left) and Carotid Artery (right). The viscoelastic model predicts a smoother waveform than the elastic model.}
\label{fig:simulated_sheep}
\end{figure}

Figure~\ref{fig:simulated_sheep} presents the simulated results of
flow rate at two different representative locations: Medial Descending
Aorta (left) en Carotid Artery (right) for peripheral arteries. The elastic wall model shows high frequency components, specially on the Carotid Artery. With the viscoelastic wall model we observe on the contrary that the high frequency components of the waveform are damped.

Previous numerical
studies~\cite{holenstein1980viscoelastic,raghu2011comparative,reymond2009validation,segers1997assessment,steele2011predicting}
have shown the significant damping effect of wall viscosity on the
pulse waves but limited by the lack of exactitude of the values of the model
parameters, especially for the viscoelasticity of the arterial
network. In our numerical simulations we use estimates of the viscoelasticity by evaluating the pressure-diameter relationship from a dataset of direct measurements on  arterial network of sheep.

One of majors the drawbacks of numerical simulation on extended networks is the impossibility of computing the viscoelastic coefficients directly from experimental data. On the contrary the Young's modulus is well known and a large literature exists. If we work with the hypothesis that the ratio between the Young's modulus is almost constant we will able to build networks using accessible information.

\section{Conclusion}

We estimated the viscoelasticity of the  arterial network of a sheep by evaluating the pressure-diameter relationship with a dataset of direct measurements. 
Good agreements between a proposed nonlinear Kelvin-Voigt model and measurements were achieved through a linear regression method.
The obtained parameter values were used in a 1D blood flow model to simulate the pulse waves in the arterial network.
We have shown the damping effect of the wall viscosity on the high frequency waves, especially at the peripheral arteries.
We explained it by the nearly constant value of the viscoelastic relaxation time,  defined by the ratio between 
the viscosity coefficient and the Young's modulus.
The optimal values of the ratio $ {\phi_{NL} \over \phi } $ seems to be constant in five of the seven arteries, we plan for a future work to study the impact of the nonlinear coefficient $\phi_{NL}$ in large networks. 
\bibliographystyle{asmems4}
\bibliography{Numerical}
\end{document}